\documentclass[%
floatfix,
showpacs,
superscriptaddress,
aps,
prc,
nofootinbib,
reprint]{revtex4-1}

\usepackage{hyperref}
\usepackage[T1]{fontenc}
\usepackage{comment}
\usepackage{tgtermes}
\usepackage{graphicx}
\usepackage{color}
\usepackage{dsfont}
\usepackage{bm}
\usepackage{slashed}
\usepackage{changepage} 
\usepackage{mathtools}
\usepackage{amsmath}
\usepackage{subfigure}
\usepackage{indentfirst}
\usepackage{hyperref}
\usepackage{booktabs, tabularx}
\usepackage{enumitem}
\usepackage[utf8]{inputenc}
\usepackage{ulem}

\hypersetup{
    colorlinks=true,
    linkcolor=blue,
    filecolor=magenta,      
    urlcolor=blue,
    citecolor=blue
}

\usepackage{amssymb}
\usepackage{multirow,array}
\usepackage{lipsum}
\usepackage{enumitem}

\newcommand{\be}{\begin{equation}}
\newcommand{\ee}{\end{equation}}
\newcommand{\bs}{\begin{subequations}}
\newcommand{\es}{\end{subequations}}
\newcommand{\beal}{\begin{align}}

\newcommand{\trento}{\texttt{T$_\mathrm{R}$ENTo}}
\newcommand{\ktiso}{\texttt{K$_\mathrm{T}$Iso}}
\newcommand{\sqrts}{\sqrt{s_\textrm{NN}}}

\newcommand{\tauiso}{\tau_{\rm iso}}
\newcommand{\etabar}{\bar{\eta}_T}

\definecolor{dartmouthgreen}{rgb}{0.05, 0.5, 0.06}

\begin{document}
\title{Pre-hydrodynamic evolution and its impact on quark-gluon plasma signatures}

\author{Dananjaya Liyanage}
\affiliation{Department of Physics, The Ohio State University, Columbus, OH 43210, USA}
\author{Derek Everett}
\affiliation{Department of Physics, The Ohio State University, Columbus, OH 43210, USA}
\author{Chandrodoy Chattopadhyay}
\affiliation{Department of Physics, The Ohio State University, Columbus, OH 43210, USA}
\affiliation{Department of Physics, North Carolina State University, Raleigh, NC 27695, USA}
\author{Ulrich Heinz}
\affiliation{Department of Physics, The Ohio State University, Columbus, OH 43210, USA}

\begin{abstract}
    State-of-the-art hydrodynamic models of heavy-ion collisions have considerable theoretical model uncertainties in the description of the very early pre-hydrodynamic stage. We add a new computational module, \ktiso{}, that describes the pre-hydrodynamic evolution kinetically, based on the relativistic Boltzmann equation with collisions treated in the Isotropization Time Approximation. As a novelty, \ktiso{} allows for the inclusion and evolution of initial-state momentum anisotropies. To maintain computational efficiency \ktiso{} assumes strict longitudinal boost invariance and allows collisions to isotropize only the transverse momenta. We use it to explore the sensitivity of hadronic observables measured in relativistic heavy-ion collisions to initial-state momentum anisotropies and microscopic scattering during the pre-hydrodynamic stage.   
\end{abstract}

\maketitle

\section{Introduction}
\label{sec1}
\vspace*{-3mm}

State-of-the-art estimations of the transport properties of Quark Gluon Plasma from heavy-ion collision experiments still include relatively large theoretical uncertainties in modeling the early stages of the collision. This theoretical uncertainty is folded into the estimate of the total error on the transport coefficients $(\eta/s)(T)$ and $(\zeta/s)(T)$ when performing statistical model-to-data comparisons, and it limits the knowledge extraction from experimental data irrespective of the precision of the latter. This provides a powerful incentive to practitioners to improve the description of the early-time dynamics, reducing potential sources of model bias. 

Recent Bayesian studies of both large and small heavy-ion collision systems \cite{ Moreland:2018jos, Bernhard:2019bmu, JETSCAPE:2020shq, Nijs:2020ors, Nijs:2020roc, Everett:2020xug} have mostly employed a pre-hydrodynamic model based on free-streaming of massless partons. In this model the initial energy deposition is parametrized by \trento{}, followed by free-streaming expansion for a brief proper time interval of $O(1)$\,fm/$c$ before the stress tensor is Landau matched to viscous hydrodynamics. An improved approach called K{\o}MP{\o}ST \cite{Kurkela:2018wud, Kurkela:2018vqr} replaces the free-streaming stage by a relativistic effective kinetic theory motivated by perturbative QCD dynamics \cite{Arnold:2002zm, Kurkela:2015qoa, Keegan:2016cpi}. It assumes the energy momentum tensor can be separated into a local average and a small perturbation around it. Then the evolution of the small perturbation around the background is treated using (to first approximation) linear response theory. One of the limitations of this realistic pre-equilibrium module is that it can not straightforwardly be applied for heavy-ion collisions with large transverse gradients in the initial energy deposition (i.e. small systems such as proton-nucleus collisions). This limitation was recently by-passed in \cite{Ambrus:2021fej} where a pre-hydrodynamic model based on kinetic theory with a simplified relaxation-type Boltzmann collisional kernel was implemented. In our work we take a similar approach but simplify the collision term further, for computational economy: unlike \cite{Ambrus:2021fej}, we allow for thermalization by rescattering of {\it only the transverse momenta} while evolving the longitudinal momenta by free-streaming with exact boost-invariance.

Our starting point is the relativistic Boltzmann equation for massless partons in the Isotropization Time Approximation (ITA) developed in Refs.~\cite{Kurkela:2018ygx, Kurkela:2019kip, Kurkela:2020wwb}. The focus of our work is on developing an efficient and flexible numerical scheme, \ktiso{} \cite{ktiso_code}, that allows to study the evolution of initial momentum anisotropies. Longitudinal boost-invariance is the price we pay for numerical efficiency and practical applicability in Bayesian model calibration; we expect this rather drastic approximation to be relaxed in future generalizations of the approach.  

\vspace*{-2mm}
\section{The Isotropization Time Approximation}
\label{sec2}
\vspace*{-2mm}

The Isotropization Time Approximation (ITA) model of the Boltzmann equation was introduced and thoroughly studied in Refs.~\cite{Kurkela:2018ygx, Kurkela:2019kip, Kurkela:2020wwb}. For a self-contained discussion we briefly summarize its main features.

\subsection{Propagation in Cartesian coordinates}
\label{sec2a}
\vspace*{-2mm}

In Cartesian coordinates, the relativistic Boltzmann equation is expressed as
\begin{equation} \label{Boltzmann}
    P^{\mu}\partial_{\mu} f(X;P) = C[f]
\end{equation}
where $P^{\mu}$ is the on-shell four-momentum vector, $X^{\mu}$ is the space-time position vector, $f(X;P)$ the one-particle distribution function, and $C[f]$ is the collision kernel. 
The authors of \cite{Kurkela:2018ygx, Kurkela:2019kip} chose to replace the full collision kernel with a simplified model which retains some of its salient features. Collisions tend to drive the system towards momentum-space isotropy in the fluid's local rest frame (LRF), so they chose a collision term given by
\begin{equation}
     C[f] = -\frac{u_{\mu} P^{\mu}}{\tauiso}\,(f{-}f_{\rm iso})
\end{equation}
where $f_{\rm iso}$ is any distribution that is isotropic (not necessarily of local equilibrium form) in the local rest frame. The flow velocity $u^{\mu}(X)$ specifies the local rest frame at point $X$ and is defined by the Landau matching condition
\begin{equation} \label{eigen}
     T^{\ \nu}_{\mu} u_{\nu} = \epsilon u_{\mu},
\end{equation}
where $\epsilon$ is the LRF energy density of the fluid. The isotropization time $\tauiso$ is the time scale on which collisions drive the system towards isotropy. 

We assume that the system is composed of massless degrees of freedom $P^{\mu}P_{\mu} = 0$ with conformal symmetry such that $E_p = |\bm{P}| \equiv p$. This assumption allows us to evolve only a single moment of the distribution function, and reconstruct from it the full energy-momentum tensor $T^{\mu\nu}$ at any time. Further, we define the particle velocity 4-vector 
\be
    v^{\mu} \equiv \frac{P^{\mu}}{n\cdot P},
\ee
where the unit 4-vector $n^\mu$ represents the time direction in the global frame. For massless particles $v^\mu$ is a null vector, $v^\mu v_\mu=0$, and in the global frame (where $n\cdot P=P^0$) $v^\mu=(1,\bm{v})$. The ITA model equation can then be written as
\begin{equation}
\label{ITABE}
    v^{\mu}\partial_{\mu} f(X;P) = -\frac{u \cdot v}{\tauiso}(f - f_{\rm iso}).
\end{equation}
Since $f_{\rm iso}$ is a Lorentz scalar, conformally symmetric, and isotropic in the fluid's local rest frame, it must be of the form 
\be
    f_{\rm iso}(X;P) = f_{\rm iso}\Bigl(P\cdot u(X) /\Lambda(X)\Bigr)
\ee
where the single energy scale $\Lambda(X)$ plays the role of an effective temperature and controls the local energy density $\epsilon(X)$ of the system. The functional form of $f_{\mathrm{iso}}$ will be left undetermined.

Following \cite{Kurkela:2018ygx, Kurkela:2019kip} we define the following moment of the distribution function ($g$ denotes the number of massless degrees of freedom): 
\begin{eqnarray} \label{Kurk_F}
    F(X;\bm{\Omega}_p) &\equiv& \frac{g}{2\pi^2} \int_0^\infty \frac{p^2dp}{E_p} \, (n \cdot P)^2 f(X;P)
\nonumber\\
    &=& \frac{g}{2\pi^2} \int_0^\infty p^3\, dp\, f(X;p,\bm{\Omega}_p).
\end{eqnarray}
The integral in the last expression is written in global frame momentum coordinates $\bm{p}=(p,\bm{\Omega}_p)$ where $n\cdot P=E_p=p$ is the particle energy and $\bm{\Omega}_p$ the spatial direction of its momentum $P^\mu$ in the global frame. The energy-momentum tensor is given by averaging this moment over the global-frame momentum-space angles:
\begin{equation} 
\label{Tmunu}
   T^{\mu\nu} (x) = \int  \frac{d^2\Omega_p}{4\pi}\, v^{\mu} v^{\nu} F(x; \bm{\Omega}_p).
\end{equation}
By taking the corresponding moment of the ITA Boltzmann equation (\ref{ITABE}) one obtains
\begin{equation}
\label{C[F]}
    v^{\mu}\partial_{\mu} F = -\frac{u \cdot v}{\tauiso}(F - F_{\rm iso})
    \equiv C[F]
\end{equation}
where $F_{\rm iso}$ is the isotropic version of the moment (\ref{Kurk_F}), again using global frame integration variables:
\begin{equation} 
\label{F_iso}
    F_{\rm iso}(X;\bm{\Omega}_p) \equiv \frac{g}{2\pi^2} \int_0^\infty dp\, p^3\, f_{\rm iso}(u\cdot P/\Lambda).
\end{equation}
Note that, even though $f_{\rm iso}(u\cdot P/\Lambda)$ is isotropic in the LRF, the moment $F_{\rm iso}$ is not isotropic in the global frame, due to the flow-boost between the local and global reference frames. Still, $F_{\rm iso}$ is uniquely determined by Landau matching to the energy density $\epsilon$ in the local fluid rest frame: 
\begin{align}
\label{epsilon}
    \epsilon &= u_\mu T^{\mu\nu} u_\nu 
    = \frac{g}{(2\pi)^3} \int \frac{d^3p}{E_p} (u\cdot p)^2 f(X;P) 
\nonumber\\
    &= \frac{g}{(2\pi)^3} \int \frac{d^3p}{E_p} (u \cdot p)^2 f_{\rm iso}(u\cdot P/\Lambda) 
     = \frac{g\mu\Lambda^4}{2\pi^2}.
\end{align}
In the last step we used LRF momentum coordinates (in which $u\cdot P$ reduces to the particle energy $E_p=p$ in the LRF) to perform the momentum integral; the value of the unitless constant $\mu =\int_0^\infty dz\, z^3\, f_{\rm iso}(z)$ distinguishes between different functional dependences (Boltzmann vs. Fermi-Dirac vs. Bose-Einstein, thermal vs. non-thermal\footnote{%
    For a local-equilibrium Boltzmann distribution $f_{\rm iso}(z){\,=\,}\exp(-z)$ with temperature $T$ given by the scale $\Lambda$, $\mu{\,=\,}3!$ such that the energy density takes the familiar form $\epsilon = 3g T^4/\pi^2$.}%
) of the locally isotropic distribution $f_{\rm iso}$ on the LRF particle energy. To relate the integral in the second line of Eq.~(\ref{epsilon}) to the moment $F_{\rm iso}(X;\bm{\Omega}_p)$ we rewrite it in the form
\begin{equation}
\nonumber
    \epsilon = \frac{g}{(2\pi)^3} \int \frac{d^3p}{E_p}  \, (u{\,\cdot\,}v)^2 (n{\,\cdot\,}P)^2\,f_{\rm iso}\left(\frac{(u{\,\cdot\,}v)(n{\,\cdot\,}P)}{\Lambda}\right)
\end{equation}
and work out the integral in global frame momentum coordinates:
\begin{align}
    \epsilon &= \int \frac{d^2\Omega_p}{4\pi}\, (u\cdot v)^2\, \frac{g}{2\pi^2}\int_0^\infty dp\,p^3\,f_{\rm iso}\left((u\cdot v)\frac{p}{\Lambda}\right)
\nonumber\\
    & = \int \frac{d^2\Omega_p}{4\pi}\, (u\cdot v)^2 \frac{g\mu}{2\pi^2}\left(\frac{\Lambda}{u\cdot v}\right)^4. 
\end{align}
Comparison with Eqs.~(\ref{F_iso}) and (\ref{epsilon}) yields the identification
\begin{align}
\label{F_iso2}
    F_{\rm iso}(X;\bm{\Omega}_p) = \frac{\epsilon}{(u{\,\cdot\,}v)^4} = \frac{u{\,\cdot\,}T{\,\cdot\,}u}{(u{\,\cdot\,}v)^4}
\end{align}
together with the useful identity
\begin{equation}
    \int \frac{d^2\Omega_p}{4\pi}\, (v\cdot u)^{-2} = 1.
\end{equation}
Eq.~(\ref{F_iso2}) shows that the moment $F_{\rm iso}$ is a Lorentz scalar. Note that the dependence of $F_{\rm iso}(X;\bm{\Omega}_p)$ in Eq.~(\ref{F_iso2}) on the angles $\bm{\Omega}_p$ of the particle velocity $\bm{v}$ is anisotropic in the global frame whenever the fluid moves with non-zero velocity in that frame. This anisotropy is generated by the factor $(v\cdot u(X))^{-4}$ on the right hand side of Eq.~(\ref{F_iso2}).

In the global frame the ITA equation of motion for the moment $F(x;\bm{\Omega}_p)$ reads
\begin{equation} \label{EoM}
    \partial_t F = -v^i \partial_i F -\frac{(u \cdot v)}{\tauiso}(F - F_{\rm iso}).
\end{equation}
Here $i=1,2,3$ runs over the spatial components. After each time step the Landau matching conditions for the new energy density $\epsilon(x)$ and flow $u^\mu(x)$ must be solved. 

For the isotropization time we adopt the conformal relation from Boltzmann kinetic theory for massless systems \cite{Romatschke:2011qp} 
\begin{equation}
\label{Ttau}
    T \tauiso = 5 \frac{\eta}{s} 
\end{equation}
where $s$ is the entropy density, $\eta/s$ the specific shear viscosity, and $T$ the temperature. The temperature is derived from the LRF energy density using the conformal equation of state,
\begin{equation}
    \epsilon(x) = a T^4(x),
\end{equation}
where $a=g\frac{3}{\pi^2}$ for massless particles with Boltzmann statistics and $a=g\frac{\pi^2}{30}$ for massless bosons, with $g$ denoting the number of massless degrees of freedom in the gas. It is useful to treat the dimensionless quantity $\eta/s$ as a model parameter which characterizes its interaction strength. 

\subsection{Propagation in Milne coordinates}
\label{sec2b}

The Milne coordinate system is useful for describing high-energy heavy-ion collisions which have approximate boost-invariance along the beam ($z$) direction. This coordinate system $x^{\mu} = (\tau, x, y, \eta)$ is related to Cartesian coordinates by the relations
\begin{equation}
    \tau = \sqrt{t^2 - z^2}; \qquad 
    \eta = \frac{1}{2}\ln\left(\frac{t+z}{t - z}\right), 
\end{equation}
where $\tau$ is the longitudinal proper time (often called just the `proper time'), and $\eta$ the space-time rapidity. 

Assuming longitudinal boost-invariance (i.e. $f$ can depend on $z$ and $p_z$ only through the boost-invariant combination $w\equiv tp_z{\,-\,}zE_p$), we can focus our attention at mid-rapidity, $\eta_s=0$, where the massless Boltzmann equation reads in Milne coordinates \cite{Baym:1984np}
\begin{equation}
    (\partial_{\tau} + v^i \partial_i - \frac{p_z}{\tau} \partial_{p_z}) f = C[f];
\end{equation}
here $i=1,2$ sums over the transverse coordinates $(x,y)$ and $p_z\equiv p^3$ is the longitudinal component of $P^\mu$. In terms of the moment $F(X;\bm{\Omega}_p)=F(X;\phi_p,v_z)$ defined above this can be rewritten as \cite{Kurkela:2018ygx, Kurkela:2019kip}
\begin{equation}
\label{EOM_Milne}
    \left(\partial_{\tau} + v^i \partial_i - \frac{1}{\tau}v_z(1-v_z^2)\partial_{v_z} + \frac{4v_z^2}{\tau}\right) F = C[F]. 
\end{equation}
This is the equation of motion for longitudinally boost-invariant massless systems in the Isotropization Time Approximation. To evaluate the collision term from Eqs.~(\ref{C[F]},\ref{F_iso2}), at each time step of the evolution the LRF energy density and flow velocity must be found from Eq.~(\ref{eigen}). This requires computing the energy-momentum tensor (\ref{Tmunu}) by numerically integrating the moment $F$ over the angular variables in momentum space. Expressing the angular integral over $d^2\Omega_p$ in terms of the azimuthal angle around the $z$-axis, $\phi_p$, and the cosine of the polar angle, $v_z$, it is given by
\begin{equation}
    T^{\mu\nu}(x) =  \int \frac{dv_z}{2}\frac{d\phi_p}{2\pi}v^{\mu} v^{\nu} F(x;\phi_p,v_z). 
\end{equation}
%

\section{Reduction to boost-invariant degrees of freedom}
\label{sec3}

In practice, performing a Bayesian parameter estimation for the type of sophisticated dynamical models presently being used to describe heavy-ion collisions (see, e.g., Refs.~\cite{ JETSCAPE:2020shq, Everett:2020xug, Nijs:2020roc, Nijs:2020ors}) with acceptable statistical uncertainty requires a model that runs no longer than about an hour per collision event. At this point, using Eq.~(\ref{EOM_Milne}) we cannot achieve this even for only the pre-hydrodynamic stage without additional approximations. For this practical reason, we simplify the numerical equations further by reducing the dimensionality of momentum space,  requiring that the distribution function at $\eta_s=0$ stay proportional to $\delta(v_z)$ throughout the pre-hydrodynamic evolution.%
\footnote{%
    This is our point of departure from \cite{Ambrus:2021fej} where the effects of all $v_z$-dependent terms on the l.h.s. of Eq.~(\ref{EOM_Milne}) and in the collisional kernel are consistently incorporated. Accordingly, the microscopic dynamics in \cite{Ambrus:2021fej} generates a non-vanishing effective longitudinal pressure even if it is initially zero, whereas in our approach $P_L$ is assumed to stay zero throughout the pre-hydrodynamic stage.}

At very early times the rapid longitudinal expansion rate drives the system toward longitudinal free-streaming, such that the effect of the collisional kernel is substantially limited to isotropizing the transverse momentum dependence of the distribution function. This may be understood from Eq.~(\ref{EOM_Milne}) as follows: On the left hand side there are two terms involving $v_z$, one term involving the derivative $\partial_{v_z}$, the other a geometric term $\propto v_z^2$. Together these two terms act to make the function $F$ sharply peaked around $v_z{\,=\,}0$. In the absence of the collision term on the r.h.s., {\it any} initial distribution for $F$ at $\eta_s=0$ thus eventually approaches a $\delta$-function peaked at zero, $\delta(v_z)$. Such a momentum distribution corresponds to vanishing effective longitudinal pressure, $P_L{\,=\,}0$, throughout the pre-hydrodynamic evolution.

At asymptotically early times, the longitudinal expansion rate is $\theta_L{\,=\,}1/\tau$ while the initial transverse expansion rate $\theta_T(\tau{=}0){\,=\,}0$ by assumption $\bigl(u^x(x, y, \tau{=}0,) = u^y(x, y,\tau{=}0) = 0\bigr)$. Therefore, at early times the longitudinal Knudsen number is much larger than the transverse Knudsen number, $\tauiso  \theta_L \gg \tauiso  \theta_T$. If we restrict the evolution following Eq.~(\ref{EOM_Milne}) to times $\tau \lesssim 1$\,fm/$c$, the condition $\tauiso \theta_L{\,\gtrsim\,}\tauiso\theta_T$ is satisfied throughout the evolution. 

Under these approximations, the equations of motion simplify. Writing $F(\tau,\bm{x}_\perp;\phi_p,v_z){\,=\,}\delta(v_z) \,\tilde{F}(\tau,\bm{x}_\perp,\phi_p)$, where $\tilde{F}(\tau,\bm{x}_\perp,\phi_p){\,=\,}\int dv_z\, F(\tau,\bm{x}_\perp;\phi_p,v_z)$, Eq.~(\ref{EOM_Milne}) implies that the reduced moment $\tilde F$ evolves according to
\begin{equation}
\label{EOM_Milne_boost_inv}
    \left(\partial_{\tau} + v^i \partial_i + \frac{1}{\tau} \right) \tilde F = C[\tilde F] 
\end{equation}
or, discretized in time, as
\begin{eqnarray}
\label{eqn:eom_milne_boost_inv}
  &&\tilde F(\tau+\Delta\tau, \bm{x}_\perp; \phi_p) 
    = 
\\\nonumber
  &&\tilde F(\tau, \bm{x}_\perp; \phi_p)
    -\Delta\tau\left[   
    v^i \partial_i \tilde F
    + \frac{\tilde F}{\tau}
    + \frac{u{\,\cdot\,}v}{\tauiso}(\tilde F{-}\tilde{F}_{\rm iso})
    \right].
\end{eqnarray}

Although this approximation prevents us from studying the interesting issue of isotropization of the longitudinal and transverse pressures, it yields a kinetic model that goes beyond free-streaming and can still be used for event-by-event simulations with fluctuating initial transverse density profiles. Such simulations are required for studying the evolution of anisotropic transverse flow in both large (nucleus-nucleus) and small (proton-proton) collision systems. To remind the reader that the approximated collision term only isotropizes the {\it transverse} momenta in the $(x, y)$-plane, we denote the specific shear viscosity $\eta/s$ in Eq.~(\ref{Ttau}) from here on by $\etabar$.

We use operator splitting to define separate time propagation operators for each of the three terms in parentheses in Eq.~(\ref{eqn:eom_milne_boost_inv}). We now discuss each of them in turn.  

\paragraph{Transverse Advection.}
The first term describes spatial advection by free-streaming in the transverse plane and is handled with the MacCormack scheme \cite{john1995computational}. For an equation of the form
\begin{equation} 
     \frac{\partial \rho}{\partial t} + a \frac{\partial \rho}{\partial x} = 0
\end{equation}
the MacCormack scheme proceeds in two steps: in the first, ``prediction'' step, derivatives are replaced by forward differences:
\begin{equation}
    \bar{\rho}_i^{n+1} = \rho_i^n - a \frac{\Delta t}{\Delta x} (\rho_{i+1}^n - \rho_i^n).
\end{equation}
Here the indices $i$ and $n$ label the spatial and temporal lattices, respectively. The second, ``correction'' step uses the time average of the prediction step and the initial value,
\begin{equation}
    \rho_i^{n+1/2} \equiv \frac{ \rho_i^{n} + \bar{\rho}_i^{n+1} }{2}, 
\end{equation}
to yield 
\begin{equation}
    \rho_i^{n+1} = \rho_i^{n+1/2} - a \frac{\Delta t}{2 \Delta x} (\bar{\rho}_{i}^{n+1} - \bar{\rho}_{i-1}^{n+1}).
\end{equation}
As a second-order flux-conserving method, the MacCormack algorithm works well to approximate the spatial advection as long as the spatial grid spacing satisfies $\Delta x < \sigma / 6$ where $\sigma$ is the smallest physical scale in the transverse plane (for instance, the width of the nucleon in \trento{} or MC-Glauber initial conditions), and the proper time spacing obeys $\Delta \tau < \Delta x / 8$. 

\paragraph{Longitudinal Expansion.}
The second term results from the longitudinal Bjorken expansion. At each time step it is integrated by using the exact solution of the equation 
\be
    dF/d\tau = -F / \tau, 
\ee
given by
\be
    \tau F(\tau) = \rm const. 
\ee

\paragraph{Collisions.}
The propagation of the collision term is more difficult, because it acts locally in space-time to isotropize the momentum-space dependence of the distribution in the local rest frame. First, the stress-tensor $T^{\mu\nu}$ is calculated by integrating over the transverse momentum azimuthal angle $\phi_p$. Then, the energy density and flow velocity follow from Landau-matching, which allows us to calculate the collision term. We have found that using the fourth-order Runge-Kutta method to propagate the collision term gives sufficient accuracy and energy-conservation as long as $\etabar$ is not too small and the temporal step-size $\Delta \tau$ is chosen to be much smaller than the isotropization time across the entire grid, $\Delta \tau \lesssim \tauiso / 8$. 

\paragraph{Adaptive time steps.}
In practice we exhaustively search for the smallest value of $\tauiso$ on the entire grid $\tauiso^{\rm min}$ at each proper time step, and set the proper time step-size for the next iteration accordingly. The proper time step-size also needs to be smaller than the spatial grid spacing $\Delta x$ in order for the free-streaming terms to be propagated with sufficient precision. So, at each iteration, one can set the proper time step-size for the following iteration according to 
\begin{equation}
    \Delta \tau = \min(\Delta x / 8, \tauiso^{\rm min} / 8)    
\end{equation}
where $\tauiso^{\rm min}$ is the minimal value of the isotropization time across the entire grid (i.e. the isotropization time in the cell with the largest energy density). 
\section{Initial-state momentum-space anisotropy}
\label{sec4}

Many commonly used pre-hydrodynamic modules in hybrid models (including \trento, \trento+FS) implement locally isotropic initial momentum distribution in the transverse plane and therefore generate momentum-space anisotropy only via spatially anisotropic collective expansion. Another kinetic theory based pre-equilbrium module, K{\o}MP{\o}ST, in principle allows for initial momentum anisotropies, but a study of the effects of such anisotropies has not yet been done \cite{Kurkela:2019kip}. The IP-Glasma model \cite{Schenke:2012wb}, on the other hand, does include local momentum anisotropies in the initial state. In sufficiently strongly coupled systems these initial momentum anisotropies are erased rapidly by microscopic interactions before transverse collective expansion becomes appreciable. But in an initially weakly coupled environment undergoing strong longitudinal expansion, such as the one studied here, they are propagated into the initial state of the hydrodynamic stage and ultimately affect the final state momentum anisotropies. Studying the experimental sensitivity to initial-state momentum anisotropies in order to identify possible experimental signatures \cite{Giacalone:2020byk} requires a model which has control over this effect. In this section, we consider a minimal extension of the \trento{} model which can provide a parametrized source of transverse momentum-space anisotropy in the initial state.

Let $T_R(\bm{x}_\perp)$ denote the transverse profile generated by the \trento{} model which is usually interpreted as $\tau_0\, \epsilon(\bm{x}_\perp)$, where $\epsilon(\bm{x}_\perp)$ is the initial energy density profile at the initial time $\tau_0$. To incorporate momentum-space anisotropy we assume that the initial moment $F$ takes the form
\begin{equation}
\label{Adef}
    F(\tau_0,\bm{x}_\perp; \phi_p) = \mathcal{N}\, T_R(\bm{x}_\perp)\, \mathcal{A}(\bm{x}_\perp, \phi_p)
\end{equation}
in the global frame, with a positive definite function $\mathcal{A}(\bm{x}_\perp, \phi_p)$.\footnote{%
    Consistent with the reduction to boost-invariant degrees of freedom in the preceding section, this parametrization accounts only for anisotropies in the transverse momentum distribution.}
$\mathcal{A}$ can be Fourier decomposed as usual in terms of initial-state anisotropic flow coefficients $\tilde{v}_n$ and flow-plane angles $\tilde{\psi}_n$:
\begin{equation}
\label{A}
    \mathcal{A}(\bm{x}_\perp, \phi_p) = 1 + 2 \sum_{n=1}^\infty \tilde{v}_n(\bm{x}_\perp) \cos\bigl[n\bigl(\phi_p{-}\tilde{\psi}_n(\bm{x}_\perp)\bigr)\bigr].
\end{equation}
This form is motivated in particular by the IP-Glasma model, specifically by the concept of `color-domains' \cite{Lappi:2015vta} -- disjoint patches of the transverse plane within which the color fields are aligned. The angles $\tilde{\psi}_n(\bm{x}_\perp)\in[0,2\pi/n)$  control the shape and orientation (in momentum space) of the patch surrounding the point $\bm{x}_\perp\equiv(x,y)$ in the transverse plane, with  $\tilde{v}_n(\bm{x}_\perp)\in[0,0.5)$ characterizing the relative strengths of different harmonic contributions. In \ktiso{} we only consider the effect of the elliptic flow coefficient $\bigl(\tilde{v}_2,\, \tilde{\psi}_2 \bigr)(\bm{x}_\perp)$ and assume all other coefficients to be zero. The elliptic flow coefficient $\tilde{v}_2$ is modeled using a Bessel-Gaussian \cite{Voloshin:2007pc} random field in the transverse plane $\bm{x}_\perp$. We realize such a random field by elevating $\tilde{v}_2$ to a complex variable and first generating Gaussian random fields with identical means $m$, variances $\sigma^2$ and transverse correlation lengths $l_{bg}$ for its real and imaginary parts, $\Re\tilde{v}_2$ and $\Im\tilde{v}_2$. By adding the simulation outputs of these two fields in quadrature and taking the square root we obtain a Bessel-Gaussian random field $\mathcal{BG}(m,\sigma,l_{bg})$ for $\tilde{v}_2=\sqrt{(\Re\tilde{v}_2)^2+(\Im\tilde{v}_2)^2}$. To ensure positivity of $\mathcal{A}$ in (\ref{A}) we cut off the tail of the Bessel-Gaussian random field by imposing an upper limit of 0.5 for its output value. For testing purposes we varied $m$ between $0$ and 0.3, and $\sigma$ between 0 and 0.05, and convinced ourselves that within these parameter ranges the cutoff procedure leads to only minimal distortions of the Bessel-Gaussian distribution.\footnote{%
    Simulation of initial flows with larger variances probably requires stochastically sampling the field $\mathcal{A}(\bm{x}_\perp, \phi_p)$ in (\ref{Adef}) directly.}  
The associated flow angle $\tilde{\psi}_2(x,y)$ is represented by a two-dimensional uniform random field\ $\mathcal{U}(l_u)$ where the parameter $l_u$ controls the transverse correlation length for the initial flow angles.

\begin{figure}[!b]
\includegraphics[width=9cm]{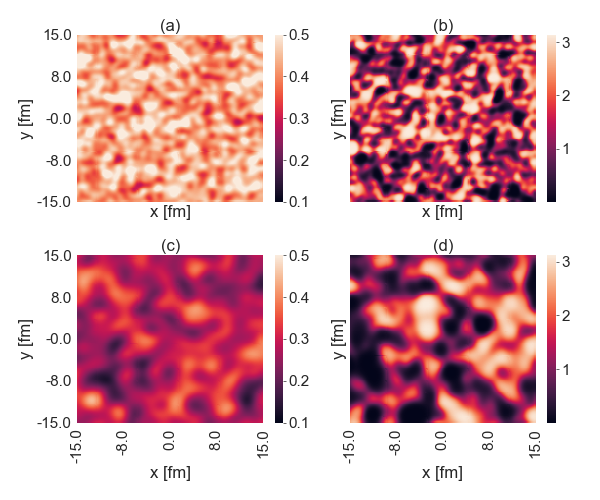}
\caption{Realizations of two dimensional random fields for $\tilde{v}_2(x,y){\,\sim\,}\mathcal{BG}(m,\sigma,l_{bg})$ (left) and $\tilde{\psi}_2(x,y){\,\sim\,}\mathcal{U}(l_u)$ (right). The upper panels (a) and (b) are generated with $m{\,=\,}0.3$, $\sigma{\,=\,}0.05$, $l_u{\,=\,}l_{bg}{\,=\,}1$\,fm. The lower panels (c), (d) are generated with $m{\,=\,}0.2, \sigma{\,=\,}0.05, l_u{\,=\,}l_{bg}{\,=\,}2$\,fm.}
\label{fig:random_fields}
\end{figure}

Fig.~\ref{fig:random_fields} shows how the mean $(m)$ and transverse correlation lengths $(l_{bg},l_{u})$ affect the structure of the random fields. The mean $m$ controls the magnitude of the initial elliptic flow in the disjoint `color-domains' in the transverse plane. Decreasing $m$ from 0.3 to 0.2 when going from (a) to (c) decreases the average value of the elliptic flow in the transverse plane. The simultaneous change in variability of $\tilde{v}_2$ between panels (a) and (c) arises from the change of the correlation lengths which control how smoothly the random fields change across the transverse plane. An increasing correlation length weakens the granularity of the random field, reducing the rate at which it varies across the transverse plane. The left panels in Fig.~\ref{fig:random_fields} illustrate this for $\tilde{v}_2$, the right panels for $\tilde{\psi}_2$, for an increase of $l_u{\,=\,}l_{bg}$ from 1\,fm in the top row to 2\,fm in the bottom row. Panels (b) and (d) illustrate how this change increases the size of the domains over which the flow angles $\tilde{\psi}_2$ are aligned with each other.

\section{Benchmarks and validations}
\label{sec5}

In this section we validate our numerical implementation, \ktiso, by comparing it to both analytic solutions and other model simulations that were used in previous studies.  

\subsection{Validation against an analytic solution}
\label{sec5a}

For the free-streaming case there exist analytic solutions for certain initial conditions, in particular for 2D transverse expansion of a longitudinally stationary system in Cartesian coordinates \cite{Romatschke:2018wgi}. Comparing with this solution provides a validation of the performance of the MacCormack algorithm for propagating the free-streaming terms in the transverse plane, as well as of our ability to reconstruct the energy density by integrating over momentum space. 

\begin{figure}[h]
\includegraphics[width=8cm]{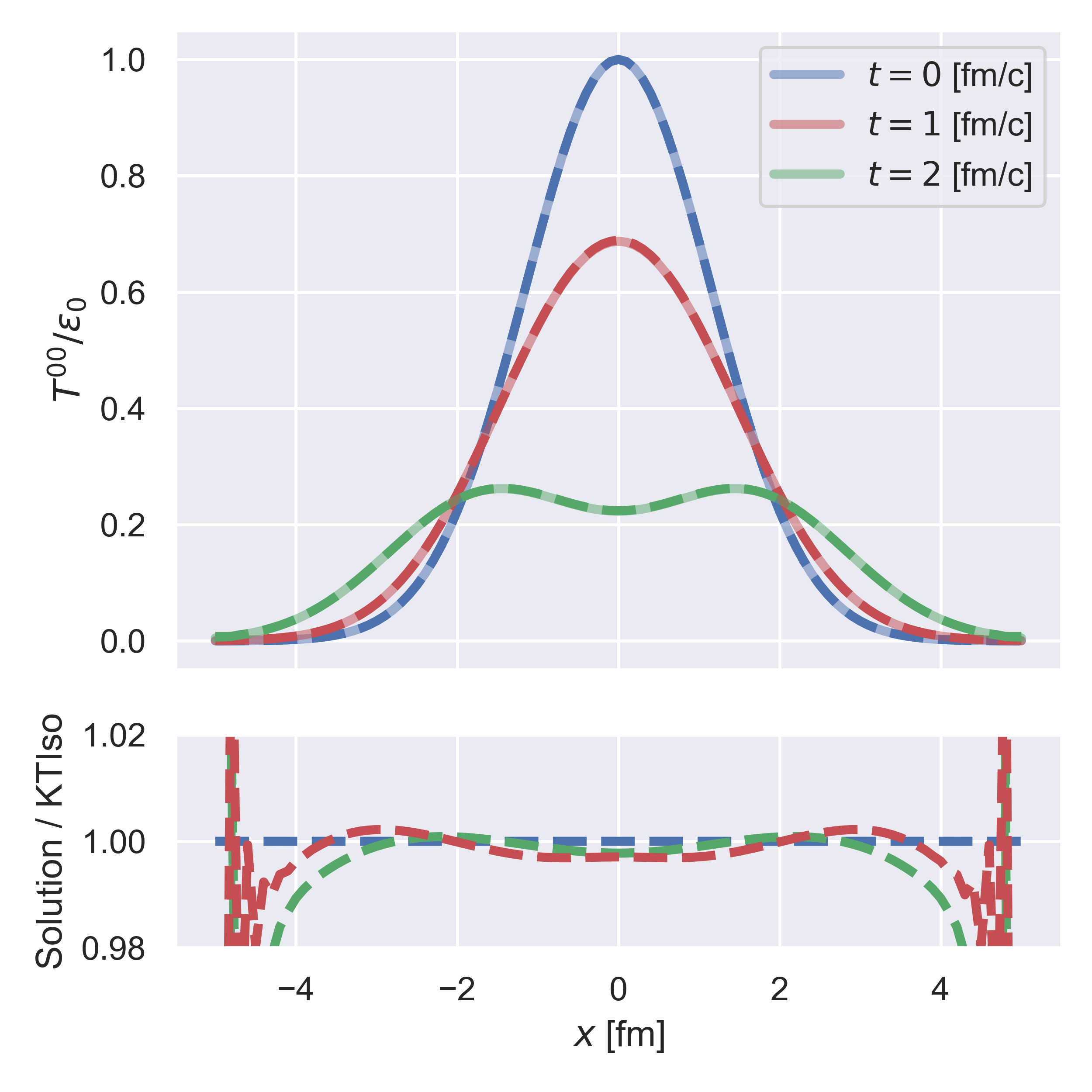}
\caption{Energy density profile in the transverse plane after free-streaming evolution. The analytic solution \cite{Romatschke:2018wgi} (solid lines) and \ktiso{} \cite{ktiso_code} implementation (dashed lines) show good agreement. The ratio of the analytic solution and \ktiso{} are shown in the lower panel. 
}
\label{validate_fs}
\end{figure}

\begin{figure*}[!htb]
\noindent\makebox[\textwidth]{%
  \centering
  \begin{minipage}{1\textwidth}
    \includegraphics[width=0.98\textwidth]{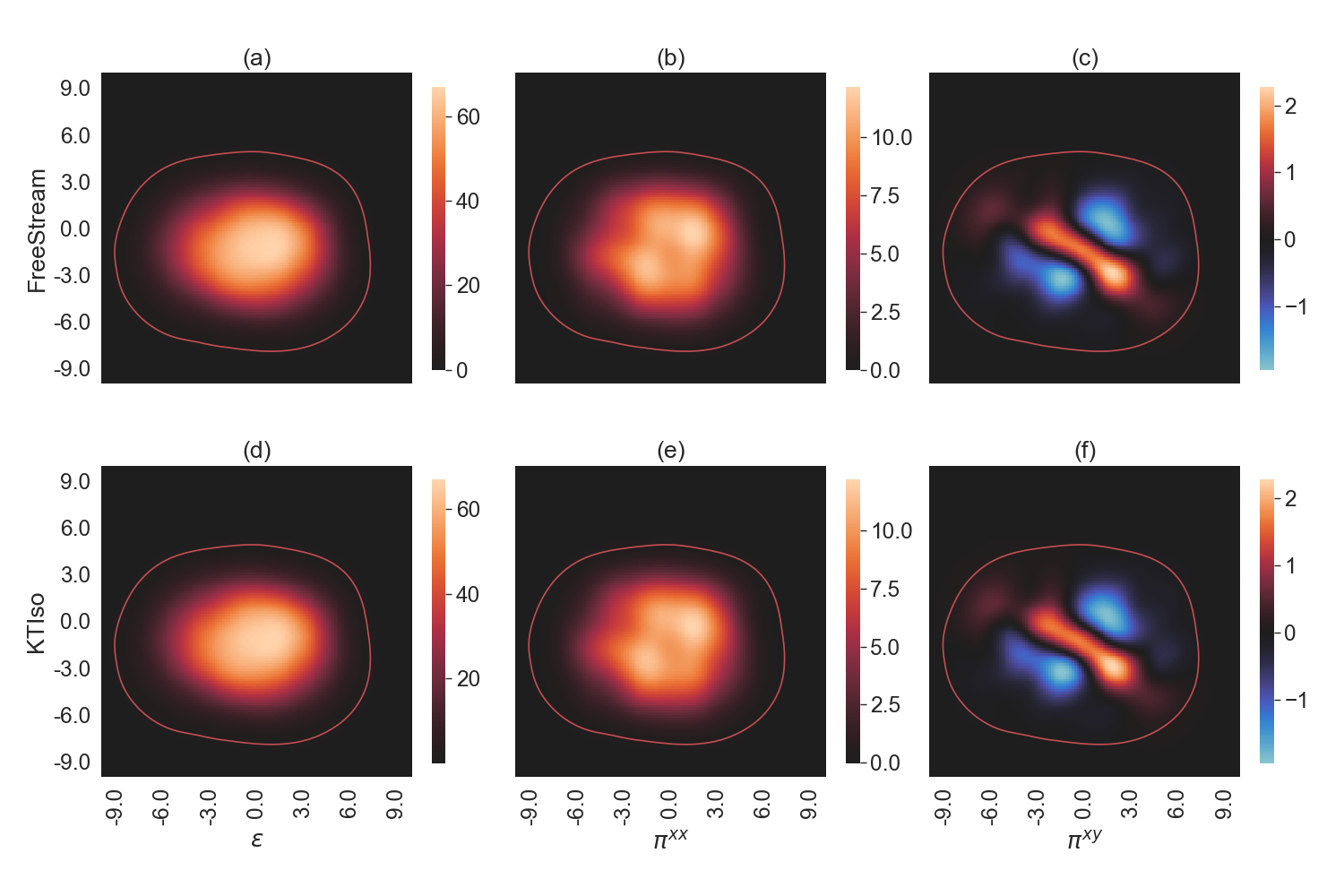}
    \vspace*{-6mm}
  \end{minipage}
  }
  \caption{
  Comparison of the results from \texttt{freestream-milne} \cite{fsmilne_code} (top row) and \ktiso{} \cite{ktiso_code} without collisions (bottom row)
  for the energy density $\epsilon$ (a,d) and the shear stress components $\pi^{xx}$ (b,e) and $\pi^{xy}$ (c,f), after $1.5$\,fm/$c$ of free-streaming evolution. The red contour denotes a surface of constant energy density $\epsilon = 1$\,GeV/fm$^3$. 
  }
  \label{fig:compare_fs_kt}
\end{figure*}
%

We follow the solution in \cite{Romatschke:2018wgi} but in two (rather than three) spatial dimensions. Moreover, the equations are propagated in Cartesian time rather than proper time, as no boost-invariant symmetry is assumed. Although this geometry is different than the usual use case, it provides a straightforward numerical check of the MacCormack algorithm's propagation of the transverse streaming terms, which are the same in either geometry. For a local equilibrium initial condition with a Gaussian spatial temperature profile in the transverse plane the distribution function at $\eta_s=0$ is initially taken as
\begin{equation}
    f_{\text{init}}(\vec{x}_\perp;p) = \pi \exp\bigl(-p/\Lambda(\vec{x}_\perp)\bigr),
\end{equation}
with
\begin{equation}
    \Lambda(\vec{x}_\perp) = T_0 \exp\left(-\frac{\vec{x}_\perp^2}{8b^2}\right).
\end{equation}
For the width $b$ of the temperature profile we take $b=1$\,fm. The initial energy density per unit area is then given by
\begin{equation}
    T_{\text{init}}^{00}(t, \vec{x}_\perp) = T_0^3 \exp\left(-\frac{3}{8}\frac{\vec{x}_\perp^2}{b^2}\right).
\end{equation}
The free-streaming solution at time $t>0$ is \cite{Romatschke:2018wgi}
\begin{equation}
    T^{00}(t, \vec{x}_\perp) = T_0^3 \exp\left(-\frac{3}{8} \frac{|\vec{x}_\perp|^2{+}t^2}{b^2}\right)
    I_0\left(-\frac{3}{4}\frac{|\vec{x}_\perp|t}{4b^2}\right),
\end{equation}
where the modified Bessel function $I_0$ arises from the azimuthal angular integral over $\phi_p$. 

In Fig.~\ref{validate_fs} we show a comparison of the LRF energy density $\epsilon$ from the numerical implementation \ktiso{} with the analytic solution at initialization and after 1 and $2$\ fm/$c$ of free-streaming evolution, showing almost perfect agreement.

\subsection{Validations against an external free-streaming code}
\label{sec5b}

We also cross-checked \ktiso{} in free-streaming mode for a fluctuating initial condition (where no analytic solutions exist) against the previously developed and validated free-streaming code \texttt{freestream-milne} \cite{fsmilne_code}. The selected initial condition represents a Pb-Pb collision at 0-5 \% centrality at the LHC (collision energy $\sqrts{} = 2.76$\,TeV) and was taken from the \trento{} model with the MAP parameters found in a recent Bayesian parameter estimation study \cite{Everett:2020xug}.

For the comparison we ran the \ktiso{} code \cite{ktiso_code} without collisions ($\etabar=\infty$) and without initial momentum anisotropy. In Fig.~\ref{fig:compare_fs_kt} we compare \texttt{freestream-milne} with \ktiso{} outputs for the LRF energy density $\epsilon$ (a,d) and the two shear stress components $\pi^{xx}$ (b,e) and $\pi^{xy}$ (c,f). We find very good agreement among all of the relevant hydrodynamic moments and no visible discrepancies. For cells inside the fireball region (indicated by the red line) the energy density and all shear stress components show better than $5\%$ agreement. This gives confidence that the algorithm in \ktiso{} captures the free-streaming dynamics of the ITA equations of motion with sufficient accuracy. 

While we have not found an exact solution against which to compare \ktiso{} when the collision term is turned on we performed standard convergence checks to convince ourselves that the code's precision does not decrease when allowing for collisions. We note (see Eq.~(\ref{eqn:eom_milne_boost_inv})) that strongly coupled systems with $\etabar\lesssim1/(4\pi)$ require a much shorter time step $\Delta\tau$ than weakly coupled ones.

\section{Breaking Conformal Symmetry}
\label{sec6}

All of the results described thus far assumed conformal symmetry. At the microscopic level, this implied that the particle degrees of freedom are massless. In Ref.~\cite{Nijs:2020roc} a single parameter was introduced to break conformal symmetry for the free-streaming evolution. The breaking of this symmetry is an essential ingredient for any pre-hydrodynamic model that is to be smoothly matched to the hydrodynamic evolution of a non-conformal QCD fluid. To ensure compatibility of \ktiso{} with Ref.~\cite{Nijs:2020roc} in the collisionless limit we include this feature as an option also in our code. Following Ref.~\cite{Nijs:2020roc} allow the user to adjust the transverse\footnote{%
    The reader is reminded that \ktiso{} assumes zero longitudinal velocities ($v_z=0$) in the local fluid rest frame (see Sec.~\ref{sec3}).}
free-streaming velocity $v_{\rm fs}=1$ for a conformal medium of massless degrees of freedom to values less than one, by rescaling the 4-velocity $v^\mu$ by a factor $v_{\rm fs}{\,<\,}1$.\footnote{%
    We note that, strictly speaking, the assumption of a single streaming velocity different from the speed of light for all particles is inconsistent with a kinetic model whose distribution function describes massive particles with a range of different momenta. Here we follow Ref.~\cite{Nijs:2020roc} in ignoring this issue for the sake of simplicity and computational economy.}
One sees easily that $v_{\rm fs}$ is naturally related to the trace of the stress-tensor: In the absence of collisions,\footnote{%
    Note that the same simple prescription does not work in the collisional case because it invalidates the derivations presented in Sec.~\ref{sec2}.}
the stress tensor at any proper time is given by the integral solution
\begin{equation}
    T^{\mu\nu}(\tau, \bm{x}_T) = \frac{\tau_0}{\tau}\int \frac{d\phi_p}{2\pi} v^{\mu}v^{\nu} T^{\tau\tau}(\tau_0, \bm{x}_T - \bm{v}_T \Delta \tau)
\end{equation}
where $\Delta \tau = \tau - \tau_0$ and $\bm{v}_T = v_{\rm fs}  \hat{\bm{p}}_T$ points in the direction of the particles'\ transverse momentum. It follows that the trace of the stress-tensor is given at any time by
\begin{equation}
    T^{\mu}_{\mu}(\tau, \bm{x}_T) = \frac{\tau_0}{\tau} (1 - v^2_{\rm fs})
    \int \frac{d\phi_p}{2\pi} T^{\tau\tau}(\tau_0, \bm{x}_T - \bm{v}_T \Delta \tau).
\end{equation}
So $v_{\rm fs}^2$ naturally controls the magnitude of conformal symmetry breaking \cite{Nijs:2020roc}.
$v_{\rm fs} = 1$ reproduces the conformal limit while $v_{\rm fs} = 0$ maximizes the trace. 

In practice, this effect is included by scaling the transverse velocities in the propagation of the  free-streaming terms as follows:
\begin{eqnarray}
\label{eqn:non_conformal_streaming}
  &&\tilde F(\tau+\Delta\tau, \vec{x}_\perp; \phi_p) 
    = 
\\\nonumber
  &&\tilde F(\tau, \vec{x}_\perp; \phi_p)
    -\Delta\tau\left[   
    v_{\rm fs} v^i \partial_i \tilde F 
    + \frac{\tilde{F}}{\tau}
    \right].
\end{eqnarray}
%

\section{Sensitivity of observables to pre-hydrodynamic transport}
\label{sec7}

To build intuition, we explore in this section the sensitivity of hadronic final-state observables to initial-state momentum anisotropies and/or the presence of collisions during the pre-hydrodynamic transport, using the modified ITA approach developed in the preceding sections within an existing hybrid-model framework. To this end we take the JETSCAPE model of Ref.~\cite{Everett:2020xug} and swap out the free-streaming module \texttt{freestream-milne} \cite{fsmilne_code} for the new \ktiso{} code, with collisions and initial momentum anisotropies turned on. This study complements a related one in Ref. \cite{NunesdaSilva:2020bfs} where the kinetic transport model K{\o}MP{\o}ST \cite{Kurkela:2018wud, Kurkela:2018vqr} is employed to describe the pre-hydrodynamic stage, albeit without initial-state momentum anisotropies. 

\begin{figure*}[!htb]
\includegraphics[width=\textwidth]{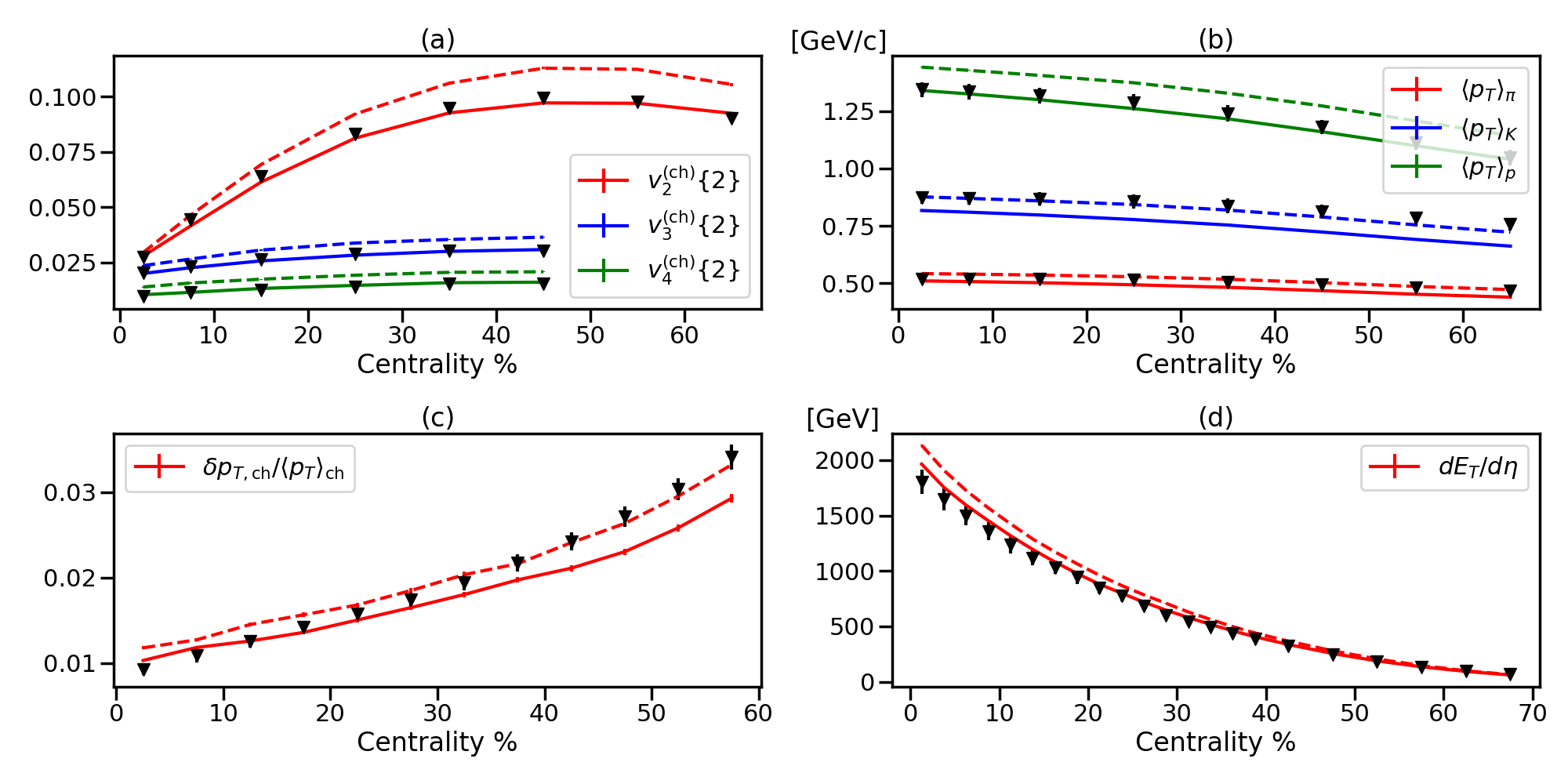}
\vspace*{-8mm}    
\caption{Observables averaged over 40,000 fluctuating initial conditions from hybrid model calculations using \ktiso{} without collisions, $\etabar = \infty$ (solid lines), and with collisions, $\etabar = 3/4\pi$ (dashed lines), shown as a function of collision centrality and compared with ALICE data for 2.76\,TeV Pb-Pb collisions \cite{ALICE:2011ab, Abelev:2013vea, Abelev:2014ckr, Adam:2016thv} (black triangles).
\vspace*{-3mm}
\label{fig:compare_visc}
}
\end{figure*}

The final-state anisotropic flow coefficients measured in nucleus-nucleus collisions are expected to arise from some combination of memory of momentum anisotropies in the initial state and hydrodynamically generated momentum anisotropies arising from spatial anisotropies in the initial-state. Both types of effects fluctuate from event to event. Previous Bayesian model parameter estimations using data from nucleus-nucleus, proton-nucleus and proton-proton collisions have employed a pre-hydrodynamic model given by free-streaming \cite{Bernhard:2015hxa, Bernhard:2016tnd, Moreland:2018gsh, Bernhard:2018hnz, Bernhard:2019bmu, JETSCAPE:2020shq, Nijs:2020ors, Nijs:2020roc, Everett:2020xug}. Including the ability for the initial momentum distribution to evolve via collisions (encoded via the parameter $\etabar$) towards local isotropy in the transverse plane is expected to render the joint estimation of the length and momentum scales characterizing the initial conditions and the shear viscosity characterizing the subsequent hydrodynamic phase more robust. In Bayesian terminology, a better estimation of both the initial energy deposition and the hydrodynamic transport is possible by marginalizing over the effects of transverse momentum isotropization due to collisions in the pre-hydrodynamic stage. 

In addition, including a parameter $\etabar$ for the shear viscosity in the pre-hydrodynamic stage could allow future Bayesian parameter estimations to constrain this parameter separately from the shear viscosity in the subsequent hydrodynamic stage. In the following we show that our model's predictions for the experimental observables exhibit sufficient sensitivity to $\etabar$ to make this a realistic possibility.

The results shown below have been obtained by running our model at a fixed set of parameters for the \trento{} initial conditions, the shear and bulk viscosities (including their temperature dependences) in the hydrodynamic phase, and the switching temperature for the particlization of the fluid in the late hadronic stage. Their values are given by the Maximum a Posteriori (MAP) parameters found in the JETSCAPE model calibration reported in \cite{Everett:2020xug} which assumed a pre-hydrodynamic stage described by free-streaming. These MAP values provide a simultaneous fit to experimental observables measured in Pb-Pb collisions at $\sqrts{} = 2.76$ TeV and in Au-Au collisions at $\sqrts{} = 0.2$ TeV. For the additional parameters in the \ktiso{} pre-equilibrium model we made judicial choices that may not be the most realistic but help to illustrate their effects on the same set of experimental observables used before, plus one additional observable that has received much recent attention \cite{Bozek:2016yoj, Giacalone:2020byk, Giacalone:2021clp}. For each set of parameters, we run 40,000 minimum bias events with fluctuating initial conditions and find centrality averages which match the experimental bins used by the ALICE experiment.

\subsection{Sensitivity to the effective shear viscosity $\etabar$ in the pre-hydrodynamics stage}
\label{sec7a}
\vspace*{-3mm}

To study the effect on final-state observables of collisions during the pre-hydrodynamic transport, we vary in the \ktiso{} module the effective shear viscosity $\etabar\equiv\eta/s$ which is related via Eq.~(\ref{Ttau}) to the isotropization time scale $\tauiso$. Specifically, we choose $\etabar=3/4\pi$ to describe a moderately strongly coupled pre-hydrodynamic stage with short isotropization time and  $\etabar=\infty$ for an extremely weakly coupled,  free-streaming pre-hydrodynamic stage (in which case we reproduce the results obtained with the \texttt{freestream-milne} module). In both cases we fix the initial momentum anisotropy to be elliptical in nature, setting the parameters of the Bessel-Gaussian random field describing these anisotropies (see Sec.~\ref{sec4}) to $m{\,=\,}0.2, \sigma{\,=\,}0.05, l_u{\,=\,}l_{bg}{\,=\,}1$\,fm (as illustrated in the upper panels of Fig.~\ref{fig:random_fields}). Interestingly, we find that the unidentified charged hadron and identified pion, kaon and proton mid-rapidity yields feature no visible sensitivity to the existence or absence of collisions during the pre-hydrodynamic stage. Among the other observables predicted by the model, only those shown in Fig.~\ref{fig:compare_visc} show any sensitivity at all to pre-hydrodynamic microscopic scattering. The solid lines show the free-streaming limit which (except for the non-zero momentum anisotropy implemented in the initial state) is the same as originally implemented in the JETSCAPE model \cite{Everett:2020xug}; the dashed lines show the changes caused by turning on pre-hydrodynamic collisions by setting $\etabar=3/4\pi$. The difference is found to be significant when compared with the precision of the experimental data from ALICE (black triangles).

\begin{figure*}[htb]
\includegraphics[width=\textwidth]{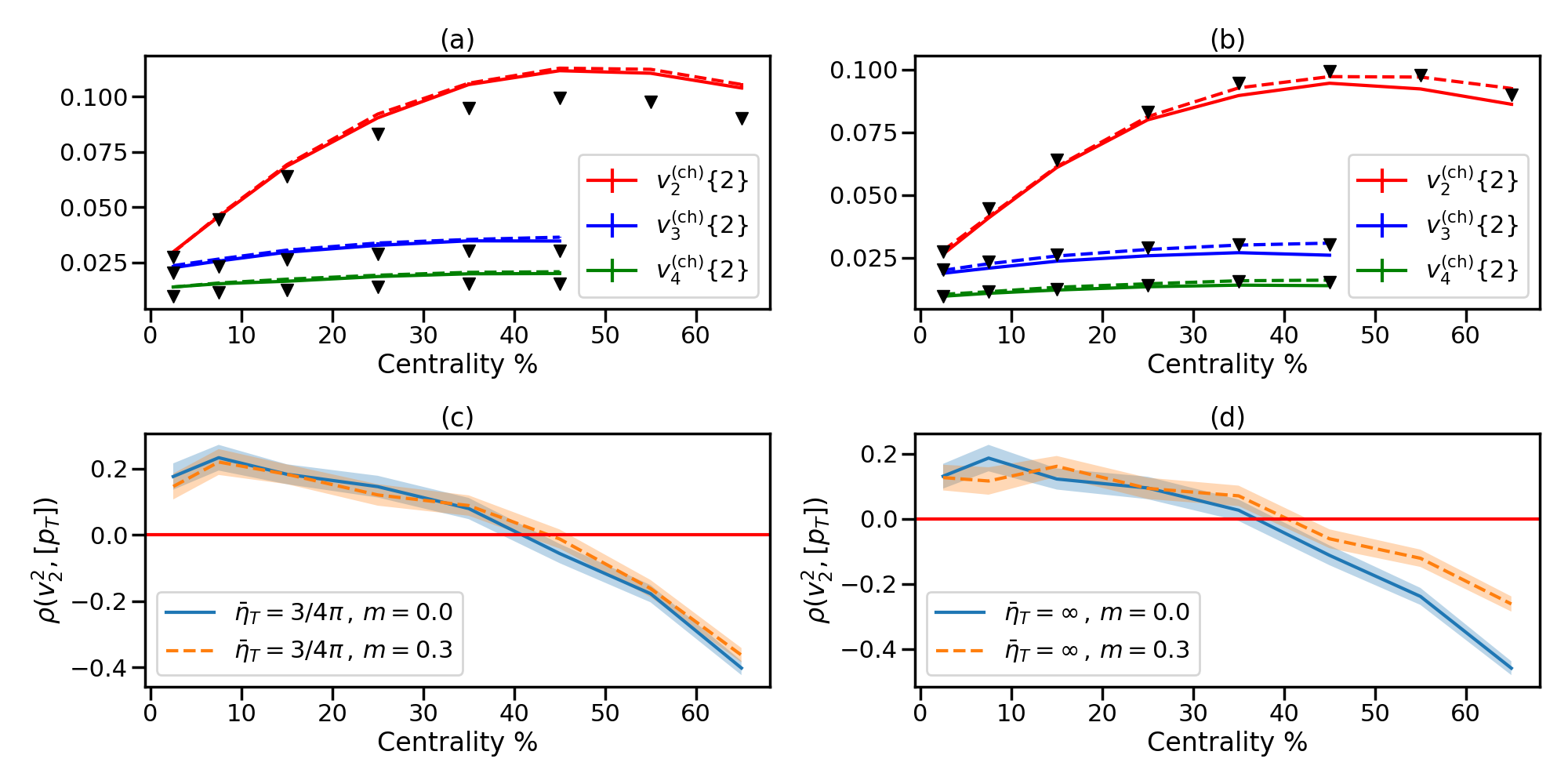}
\vspace*{-6mm}
\caption{Observables averaged over 40,000 fluctuating initial conditions from hybrid model calculations using \ktiso{} with collisions ($\etabar = 3/4\pi$, left panels) and without collisions ($\etabar = \infty$, right panels). The solid lines are for simulations with locally isotropic initial momentum distributions whereas the dashed lines are for locally anisotropic initial momentum distributions. The upper panels show flow observables from simulations as a function of collision centrality, compared with ALICE data for 2.76\,TeV Pb-Pb collisions \cite{ALICE:2011ab} (black triangles). The lower panels show the Pearson correlation coefficient between the squared two-particle cumulant elliptic flow, $(v_2\{2\})^2$, and the average transverse momentum in an event, $[p_T]$, again as a function of collision centrality. 
\vspace*{-3mm}
\label{fig:compare_mom_aniso}
}
\end{figure*}

It is well known \cite{Liu:2015nwa} that, for spatially anisotropic initial density profiles, even a free-streaming pre-hydrodynamic stage results in non-vanishing radial and anisotropic transverse flows at the beginning of the hydrodynamic stage, after matching the kinetically evolved energy-momentum tensor to hydrodynamic form. In fact, the radial transverse flow built up during free-streaming is {\it larger} than if that same stage had been described hydrodynamically, because free-streaming massless particles move outward with the speed of light but are slowed down when suffering collisions. On the other hand, faster transverse growth at early times leads to weaker pressure gradients and, correspondingly, slower growth of the radial flow velocity at later times. Overall, varying the effective shear viscosity in the early pre-hydrodynamic stage changes the balance between radial flow created at early and at later times, and the overall effect on the final-state radial flow is hard to predict without numerical simulation. Fig.~\ref{fig:compare_visc} tells us that turning on additional microscopic collisions during the pre-hydrodynamic stage leads in the final state to an overall {\it increase} of the radial flow (reflected in the mean $\langle p_T\rangle$ for pions, kaons and protons shown in panel (b)), of the anisotropic flow coefficients $v_{2,3,4}$ in panel (a), of the charged hadron $p_T$ fluctuations shown in panel (c), as well as of the overall viscous heating (reflected in an increase of the transverse energy density at midrapidity, $dE_T/d\eta$, in panel (d)), in all cases by a few percent. 

\subsection{Sensitivity to the initial momentum anisotropy in the pre-hydrodynamics stage}
\label{sec7b}
\vspace*{-3mm}

In this subsection we offer an exploratory study of observable final state effects caused by local anisotropies in the initial momentum distributions. Using the parametrization introduced in Sec.~\ref{sec4}, we control the initial momentum anisotropy by tuning the mean ($m$), variance ($\sigma^2$) and transverse correlation lengths ($l_u$ and $l_{bg})$ of the Bessel-Gaussian random fields for the initial elliptic flow vector. Fig.~\ref{fig:compare_mom_aniso} shows results for the choices
$m{\,=\,}0$ (isotropic initial momentum distribution) and $m=0.3$ (anisotropic initial momentum distribution), keeping $\sigma^2{\,=\,}0.05$ and $l_{bg}{\,=\,}1$\,fm fixed. We find that in the 2.76\,TeV Pb-Pb collisions studied here most of the initial momentum anisotropy effects are completely washed out by the hydrodynamic evolution. Specifically, the charged and identified particle yields, transverse energy, mean transverse momenta, and transverse momentum fluctuations at mid rapidity are found to have negligible sensitivity to the initial momentum anisotropy imposed in the case $m=0.3$. Weak sensitivity to $m$ is seen in the anisotropic flow coefficients; the strongest memory of a nonzero initial momentum anisotropy is seen in the most peripheral collisions, whose fireballs are small and whose hydrodynamic stage is short, when the pre-hydrodynamic stage is free-streaming ($\etabar=\infty$, panel(b)). The introduction of even a moderate amount of collisions during the pre-hydrodynamic stage ($\etabar=3/4\pi$, panel (a)) almost completely erases this memory. 

The bottom row of Fig.~\ref{fig:compare_mom_aniso} explores the effect of initial momentum anisotropies on the correlation between the square of the 2-particle cumulant elliptic flow $v_2\{2\}$ and the mean transverse momentum $[p_T]$ in an event. Pioneering work by Bo\.{z}ek found this correlator to be sensitive to the initial state of relativistic heavy-ion collisions \cite{Bozek:2016yoj}. At a fixed collision centrality, the mean transverse momentum $[p_T]$ in an event gives us a handle on the initial transverse size of the produced fireball, with larger $[p_T]$ corresponding to smaller initial radii. More compact initial configurations lead to initially larger pressure gradients, naturally driving stronger anisotropic flows. This  explains the positive sign of the correlation coefficient $\rho(v_2^2,[p_T])$ for central collisions. In peripheral collisions this relationship is complicated by the appearance of only a few, initially well-separated hot spots in the collision zone which then evolve approximately independently. In all hybrid models, except for IP-Glasma initial conditions the correlator $\rho(v_2^2,[p_T])$ was found to change sign between central and peripheral collisions \cite{Giacalone:2020byk, Giacalone:2021clp}, contrary to experimental data \cite{ATLAS:2019pvn}. Initially it was suspected that the absence of such a sign change for IP-Glasma initial conditions might be caused by initial momentum anisotropies encoded in the IP-Glasma model that were absent in all other initialization models \cite{Giacalone:2020byk}. A later study by the same authors \cite{Giacalone:2021clp} showed instead a dominant sensitivity of $\rho(v_2^2,[p_T])$ to differences in the fluctuating geometric shapes of the nucleon between the IP-Glasma and other initial-state models. Figs.~\ref{fig:compare_mom_aniso} c,d confirm the innocence of initial momentum anisotropies: while for free-streaming pre-hydrodynamic evolution (panel (d)) there is a visible effect of the initial momentum anisotropy on this correlator in peripheral collisions (where the hydrodynamic stage is brief and unable to fully erase the memory of momentum-anisotropic initialization), this effect is not large enough to avoid the sign change of $\rho(v_2^2,[p_T])$ between central and peripheral collisions, and adding even a moderate amount of scattering to the pre-hydrodynamic stage (panel (c)) completely erases this residual initial-state memory. This observation excludes initial-state momentum anisotropy in the IP-Glasma model as the culprit for the absence of a sign change for $\rho(v_2^2,[p_T])$ in IP-Glasma-initialized simulations. 

\vspace*{-3mm}
\section{Conclusions}
\label{sec8}
\vspace*{-2mm}

In this work we added a new module to the existing suite of kinetic + hydrodynamic evolution models for relativistic heavy-ion collisions, the novel pre-equilibrium dynamical model \ktiso{}. \ktiso{} is derived from the relativistic Boltzmann equation for massless particles with longitudinal boost invariance and a collision term written in Isotropization Time Approximation (ITA). It takes an intermediate spot between free-streaming (no collisions at all) and the K{\o}MP{\o}ST model \cite{Kurkela:2018wud, Kurkela:2018vqr} (in which collisions change all three spatial momentum components), by isotropizing only the transverse momentum components while free-streaming the initially boost-invariant longitudinal momenta. Setting \ktiso{} apart from most other pre-hydrodynamic models is the feature that it allows for the inclusion of event-by-event fluctuating initial-state momentum anisotropies and handles their evolution, with limited numerical overhead, through the entire pre-hydrodynamic (kinetic) stage. Different from linear response based theories such as K{\o}MP{\o}ST which use a propagator to evolve the energy-momentum tensor from an initial time to the hydrodynamic initialization time, \ktiso{} runs as a time-stepped evolution and can thus be easily extended to accommodate dynamical source terms (e.g. from jet energy loss to the QGP medium). This makes \ktiso{} a strong candidate for use as a pre-equilibrium module in future jet studies of the QGP.

While more comprehensive investigations of the phenomenological consequences of these new features for relativistic heavy-ion collisions still await execution, we here reported first results from an exploratory ``intuition-building'' study of the sensitivity to details of the pre-hydrodynmic stage of the hadronic final-state observables used in recent large-scale Bayesian model calibrations. We included in this study the correlation coefficient $\rho(v_2^2,[p_T])$ between elliptic flow and the mean transverse momentum which has attracted much recent interest. We find that the harmonic flow coefficients, mean transverse momentum, transverse energy and transverse momentum fluctuations all exhibit sufficient sensitivity to the pre-hydrodynamic effective viscosity $\etabar$ to imagine that the collisional relaxation time during the pre-hydrodynamic stage could be meaningfully constrained in future Bayesian model calibration campaigns. On the other hand, it looks very difficult if not impossible to constrain initial-state momentum anisotropies using measurements from collisions between large nuclei such as those studied here, except perhaps at very peripheral collision centralities. Especially when collisions happen already during the pre-hydrodynamic stage we observe that very few traces of any initial-momentum anisotropy survive into the final state. Small collision systems, with a shorter duration of the hydrodynamic stage, might be more useful here as they are expected to be less efficient in erasing all memory of the initial-state momentum distributions.

The correlation coefficient $\rho(v_2^2,[p_T])$ was also found to exhibit surprisingly weak sensitivity to the initial momentum anisotropy, especially if the pre-hydrodynamic evolution is strongly affected by microscopic collisions. For an extremely weakly coupled pre-hydrodynamic stage that evolves essentially by free-streaming, initial-state momentum anisotropies were seen to somewhat increase this correlator (i.e. render it slightly less negative) in very peripheral collisions.

\vspace*{-3mm}
\section*{Acknowledgements}
\vspace*{-2mm}

We thank Andi Mankolli for providing us with the code to calculate $\rho(v_2^2,[p_T])$. We gratefully acknowledge fruitful discussions with Jean-Fran\c{c}ois Paquet, Lipei Du, Michael McNelis and Matthew Luzum. This work was supported by NSF CSSI program under grant OAC-2004601 and by the National Science Foundation (NSF) within the framework of the JETSCAPE Collaboration under Award No. ACI-1550223. Additional partial support by the U.S. Department of Energy (DOE), Office of Science, Office for Nuclear Physics under Award No. DE-SC0004286 and within the framework of the BEST and JET Collaborations is also acknowledged. C.C. acknowledges support by the U.S. Department of Energy, Office of Science, Office of Nuclear Physics through the grant DE-FG02-03ER41260. U.H. would like to acknowledge support by the Alexander von Humboldt Foundation through a Humboldt Research Award, and thanks the Institut f\"ur Theoretische Physik at the J.W. Goethe-Universit\"at in Frankfurt am Main for their hospitality.

\bibliography{biblio}

\end{document}